\documentclass[conference]{IEEEtran}
\usepackage{amsmath,amsfonts,amssymb,graphicx,booktabs}
\usepackage{bm} 
\usepackage{cite}

\usepackage{algorithm}
\usepackage{makecell} 
\usepackage{algpseudocode}
\title{Dynamic Hybrid Resource Utilisation and
MCS-based Intelligent Layering }

\author{\IEEEauthorblockN{Dhrumil Bhatt}\IEEEauthorblockA{\textit{Department of Electrical and Electronics Engineering}\\ \textit{Manipal Institute of Technology}\\ \textit{Manipal Academy of Higher Education}\\ Manipal, India \\ dhrumil.bhatt@gmail.com}}

\begin{document}

\maketitle

\begin{abstract}
The coexistence of heterogeneous service classes in 5G Enhanced Mobile Broadband (eMBB), Ultra-Reliable Low Latency Communication (URLLC), and Massive Machine-Type Communication (mMTC) poses major challenges for meeting diverse Quality-of-Service (QoS) requirements under limited spectrum and power resources. Existing radio access network (RAN) slicing schemes typically optimise isolated layers or objectives, lacking physical-layer realism, slot-level adaptability, and interpretable per-slice performance metrics. This paper presents a joint optimisation framework that integrates Dynamic Hybrid Resource Utilisation with MCS-Based Intelligent Layering, formulated as a mixed-integer linear program (MILP) that jointly allocates bandwidth, power, and modulation and coding scheme (MCS) indices per slice. The model incorporates finite blocklength effects, channel misreporting, and correlated fading to ensure realistic operation. Two modes are implemented: a Baseline Mode that ensures resource-efficient QoS feasibility, and an Ideal-Chaser Mode that minimises deviation from ideal per-slice rates. Simulation results show that the proposed approach achieves energy efficiencies above $10^7$~kb/J in Baseline Mode and sub-millisecond latency with near-ideal throughput in Ideal-Chaser Mode, outperforming recent optimisation and learning-based methods in delay, fairness, and reliability. The framework provides a unified, interpretable, and computationally tractable solution for dynamic cross-layer resource management in 5G and beyond networks.
\end{abstract}


\section{Introduction}

The emergence of 5G and beyond networks heralds a new era of service heterogeneity spanning enhanced Mobile Broadband (eMBB)~\cite{itu2020,survey_b5g_2022}, Ultra-Reliable Low-Latency Communication (URLLC)~\cite{popovski2017,bennis2018}, and massive Machine-Type Communication (mMTC)~\cite{antenova2021,telit2022}. Each service class imposes distinct Quality-of-Service (QoS) requirements high throughput and efficiency for eMBB, ultra-low latency and reliability for URLLC, and scalable connectivity under power constraints for mMTC. These conflicting demands make radio access network (RAN) slicing and resource orchestration a complex multi-objective optimisation challenge, particularly under real-world constraints such as limited spectral and power resources, time-varying channels, and stringent latency requirements.

Recent studies underscore the limitations of existing RAN slicing and scheduling approaches. Joint eMBB URLLC slicing schemes that rely on puncturing or dynamic resource reallocation often struggle to preserve guaranteed eMBB rates under bursty URLLC traffic~\cite{alali2023resource}. Similarly, methods integrating modulation and coding scheme (MCS) selection with spectrum allocation remain underdeveloped in scenarios involving coupled delay–reliability constraints~\cite{gao2022hybrid}. Multi-objective optimisation techniques have been proposed to balance latency, throughput, and cost~\cite{pramanik2024cost}, yet many neglect finite blocklength effects, correlated traffic dynamics, or per-slot physical-layer feasibility. Although survey works highlight the potential of mixed-integer linear programming (MILP) in RAN resource allocation~\cite{ejaz2025survey}, most formulations lack real-time adaptability and interpretable slice-level performance indicators.

To address these challenges, this paper proposes a Dynamic Hybrid Resource Utilisation and MCS-based Intelligent Layering (DHRI) framework, a slot-level MILP model for the joint allocation of bandwidth, power, and MCS indices under realistic physical-layer conditions, including finite blocklength shaping, channel misreporting, and temporally correlated fading. The framework operates in two regimes: (i) a baseline mode that ensures strict QoS satisfaction with minimal resource usage, and (ii) an ideal chaser mode that tracks target per-slice rates by leveraging full bandwidth and power budgets.

Unlike deep reinforcement learning (DRL) W-based approaches, which rely on stochastic policy exploration and may violate hard QoS or latency constraints during training, the proposed MILP formulation offers deterministic optimality, interpretability, and feasibility guarantees at each optimisation instance. This ensures predictable and transparent control behaviour, which is essential for latency-critical or reliability-sensitive network slices.

System performance is evaluated through Monte Carlo simulations under 5G like conditions, measuring key performance indicators (KPIs) such as feasibility rate, bandwidth and power utilisation, average latency, task completion ratio (TCR), energy efficiency, fairness, and timing accuracy via the Cramér–Rao Lower Bound (CRLB). The results demonstrate that the proposed dual-mode MILP framework achieves superior delay, fairness, and energy efficiency compared to state-of-the-art optimisation and DRL-based RAN slicing methods.

\section{Literature Review}

The coexistence of heterogeneous services, such as eMBB, URLLC, and mMTC, in 5G radio access networks (RANs) necessitates slicing frameworks that can dynamically adapt to diverse QoS requirements. Conventional approaches have explored channel-aware scheduling, latency–cost optimisation, deep reinforcement learning (DRL) strategies, and mathematical optimisation models. However, most existing methods fail to achieve a unified formulation that simultaneously incorporates realistic physical-layer dynamics, per-slot MCS adaptation, and interpretable slice-level key performance indicators (KPIs).

Chen \textit{et al.}~\cite{chen2023radiosaber} proposed \emph{RadioSaber}, a two-tier channel-aware scheduler combining inter-slice and intra-slice resource allocation, achieving throughput gains of 17–72\% while maintaining fairness. Nevertheless, the model omits modulation and coding scheme (MCS) adaptation and does not incorporate finite blocklength effects. Pramanik \textit{et al.}~\cite{pramanik2024cers} introduced CERS, a multi-objective optimiser that balances latency and cost, but it lacks fading dynamics and real-time MCS adaptation. Ejaz and Choudhury~\cite{ejaz2025milpsurvey} surveyed linear, integer, and mixed-integer programming solutions for RAN resource allocation, emphasising tractability issues and limited physical-layer realism.

Korrai \textit{et al.}~\cite{korrai2023energyefficient} investigated mmWave RAN slicing via fractional programming for beam, resource, and power optimisation, but the framework does not include KPI-level tracking or slice-specific MCS adaptation. Grings \textit{et al.}~\cite{grings2025nasp} presented NASP, a hierarchical orchestration platform that reduces URLLC session setup latency by 93\%, yet lacks link-layer scheduling features such as per-slice MCS adaptation.

In the area of learning-based methods, Sun \textit{et al.}~\cite{sun2024hierarchical} proposed a hierarchical DRL framework for multi-service slicing. While effective in adaptive decision-making, the model does not capture slot-level physical-layer adaptation or finite blocklength effects. Similarly, Cai \textit{et al.}~\cite{cai2023pwdr1} proposed PW-DRL, an online predictive weighted DRL framework for network slicing, yet it omits explicit modelling of physical-layer uncertainties such as channel misreporting or short-packet reliability.

At the application layer, Balasingam \textit{et al.}~\cite{balasingam2024zipper} introduced Zipper, a predictive control paradigm offering per-application throughput and latency assurances with significantly lower tail latency. However, the model lacks PHY-layer granularity and per-slot adaptation. Eskandari \textit{et al.}~\cite{eskandari2025drl} developed an O-RAN aligned DRL framework addressing URLLC/eMBB trade-offs under service-level agreement (SLA) constraints, but did not include per-slice bandwidth or MCS selection. Lorincz \textit{et al.}~\cite{lorincz2024energy} surveyed energy-efficiency KPIs across 5G slices, providing valuable benchmarking metrics, though without coupling to dynamic physical-layer resource management.

Despite these contributions, several key limitations remain in the literature:

\begin{itemize}
  \item Absence of \textit{per-slot joint optimisation} for bandwidth, power, and MCS under realistic physical-layer constraints such as finite blocklength effects, channel misreporting, and correlated fading.
  \item Lack of \textit{dual operational modes} that differentiate between resource-efficient feasibility and target-tracking optimisation.
  \item Limited availability of frameworks that produce interpretable slice-level KPIs, including delay, task completion ratio, energy efficiency, fairness, and timing accuracy based on the Cramér–Rao lower bound (CRLB).
  \item Scarcity of \textit{MILP-based designs} that remain computationally tractable and real-time feasible for MATLAB-scale implementation.
\end{itemize}

To address the identified limitations in existing 5G network slicing approaches, this work proposes Dynamic Hybrid Resource Utilisation and MCS-based Intelligent Layering, a slot-level mixed-integer linear programming (MILP) framework that jointly optimises bandwidth, power, and MCS selection under realistic physical-layer conditions. The principal contributions and differentiating features of this study are summarised as follows:

\begin{itemize}
    \item Unified slot-level MILP formulation:  
    The proposed framework formulates a single optimisation problem that simultaneously handles power, bandwidth, and MCS decisions per time slot. This integration ensures coherent decision-making across the physical and MAC layers, enabling real-time adaptation under dynamic channel and traffic conditions.
    
    \item Realistic physical-layer integration:  
    In contrast to many existing slicing schemes that rely on simplified channel abstractions, the framework incorporates finite blocklength effects, correlated fading, and CQI-based MCS adaptation directly into the MILP constraints. This integration bridges the gap between analytical tractability and physical-layer fidelity, improving accuracy under URLLC and mixed-service scenarios.
    
    \item Two-phase lexicographic optimisation:
    A novel dual-regime control structure is introduced. The \textit{Baseline} mode guarantees strict QoS feasibility and resource efficiency, while the \textit{Ideal Chaser} mode minimises deviation from ideal slice rates without violating feasibility. This lexicographic design provides deterministic priority enforcement and prevents resource wastage, unlike stochastic convergence methods in DRL frameworks.
    
    \item Deterministic feasibility and interpretability:
    Unlike deep reinforcement learning or heuristic-based approaches that depend on empirical convergence, the proposed MILP yields deterministic solutions with explicit constraint satisfaction. This property ensures predictable latency, stable QoS guarantees, and explainable decision logic suitable for real-time network control.
    
    \item Comprehensive slice-level performance analysis:
    The system produces interpretable key performance indicators (KPIs) including delay, task completion ratio, fairness, energy efficiency, and timing accuracy (Cramér–Rao lower bound). These metrics provide fine-grained insight into the trade-offs between efficiency and service reliability across eMBB, URLLC, and mMTC slices.
    
    \item Practical and computationally efficient design:
    The per-slot MILP remains tractable for real-time RAN implementation, solving within sub-second timescales using standard solvers. This contrasts with high-complexity learning-based methods that require extensive offline training or model retraining for new channel conditions.
\end{itemize}

In summary, the proposed Dynamic Hybrid Resource Utilisation and MCS-based Intelligent Layering framework differs from existing optimisation and DRL-based approaches by offering a unified, physically accurate, and analytically interpretable control structure. It ensures deterministic QoS compliance, supports both resource-frugal and ideal-tracking operation, and provides a transparent mechanism to balance efficiency, fairness, and performance, an aspect largely unexplored in current 5G RAN slicing literature.
\section{Methodology}

This section presents the proposed slot-level mixed-integer linear programming (MILP) framework that enables dynamic, physically grounded resource allocation across heterogeneous 5G network slices. The framework jointly optimises power, bandwidth, and modulation and coding scheme (MCS) selection under realistic channel and traffic conditions, integrating both deterministic feasibility and performance-tracking objectives within a unified optimisation structure. The motivation for adopting a MILP approach arises from the need for interpretability and guaranteed constraint satisfaction, which are often unattainable in heuristic or deep reinforcement learning (DRL) methods that rely on stochastic convergence and may violate hard QoS or latency constraints during training. The proposed formulation preserves analytical rigour and yields predictable, repeatable outcomes that are essential for ultra-reliable low-latency communication (URLLC) and critical RAN control.

The system operates over a sequence of time slots, each representing a decision epoch in a Monte Carlo simulation environment characterised by correlated fading and bursty service demand. The considered network consists of $S=3$ slices enhanced Mobile Broadband (eMBB), URLLC, and massive Machine-Type Communication (mMTC) and $M=15$ standardised NR CQI indices. For each slice $s \in \{1,2,\dots, S\}$, the instantaneous channel gain $h_s > 0$ (in linear scale) evolves according to a first-order autoregressive (AR-1) process in dB to model temporally correlated fading. Each slice is subject to resource budgets $(B_{\mathrm{cap},s}, P_{\mathrm{cap},s})$, denoting the maximum available bandwidth (in PRBs) and power (in Watts) per slot. Binary variables $z_{s,m} \in \{0,1\}$ indicate CQI activation, with the exclusivity constraint $\sum_{m=1}^{M} z_{s,m} \le 1$. Continuous decision variables $B_{s,m} \ge 0$ and $P_s \ge 0$ represent the allocated bandwidth and power, respectively. The aggregate rate for slice $s$ is expressed as

\begin{equation}
R_s = \sum_{m=1}^{M} \mathrm{SE}^{\mathrm{eff}}_{s,m}\, B_{s,m},
\label{eq:Rs}
\end{equation}

where $\mathrm{SE}^{\mathrm{eff}}_{s,m}$ is the effective spectral efficiency corresponding to CQI $m$. For URLLC traffic, which operates in the finite blocklength (FBL) regime, the achievable rate is lower than the Shannon capacity. The correction is applied using the normal approximation
\begin{subequations}
\begin{align}
\mathrm{SE}^{\mathrm{eff}}(\gamma) &= \log_2(1+\gamma) - \frac{\sqrt{V(\gamma)/n}}{\ln 2}\, Q^{-1}(\varepsilon), \label{eq:FBL_SE}\\
V(\gamma) &= \left( \frac{\gamma}{1+\gamma} \right)^2, \label{eq:FBL_V}
\end{align}
\end{subequations}
where $\gamma$ is the received SNR, $n$ denotes the blocklength, and $\varepsilon$ is the target packet error probability. The term $V(\gamma)$ quantifies channel dispersion, and the second term in \eqref{eq:FBL_SE} captures the reliability penalty imposed by short-packet transmission. This treatment ensures that latency-sensitive URLLC services are modelled realistically within the same mathematical framework as eMBB and mMTC.

To preserve SINR feasibility, transmit power and bandwidth allocations are coupled through linear constraints derived from NR link-budget relations. For each $(s,m)$ pair, the required per-PRB power slope is computed as
\begin{subequations}
\begin{align}
\alpha_{s,m} &= \frac{G_{\mathrm{req},m} N_0}{h_s}, \label{eq:alpha_main}\\
G_{\mathrm{req},m} &= \Gamma_m \cdot 10^{\mathrm{NF}/10} \cdot 10^{\mathrm{IM}/10} \cdot \gamma_{\mathrm{mis}}, \label{eq:alpha_Greq}
\end{align}
\end{subequations}
where $\Gamma_m$ is the SINR threshold for CQI $m$, $N_0$ denotes the thermal noise power per PRB, $\mathrm{NF}$ and $\mathrm{IM}$ are the receiver noise figure and interference margin (in dB), and $\gamma_{\mathrm{mis}}\ge 1$ represents the expected inflation due to CQI misreporting. This formulation captures channel-awareness explicitly, ensuring that power allocation scales inversely with instantaneous gain. The linear constraints enforcing SINR and power spectral density (PSD) compliance are expressed as
\begin{equation}
P_s \ge \sum_{m=1}^{M} \alpha_{s,m} B_{s,m}, \qquad
P_s \ge \mathrm{PSD}_{\min}\sum_{m=1}^{M} B_{s,m},
\label{eq:sinr_psd}
\end{equation}
where the second term prevents unrealistically low PSD operation. Any CQIs violating per-slice PSD ceilings are masked a priori by setting $z_{s,m}=0$ and $B_{s,m}=0$ if $\alpha_{s,m} > P_{\mathrm{cap},s}/B_{\mathrm{cap},s}$.

Each slice must also satisfy a minimum per-slot throughput requirement $R_{\min,s}$, ensuring QoS compliance across services:
\begin{equation}
\sum_{m=1}^{M} \mathrm{SE}^{\mathrm{eff}}_{s,m} B_{s,m} \ge R_{\min,s}, \quad s=1,\dots,S.
\label{eq:rmin}
\end{equation}
Resource budgets are imposed as $\sum_m B_{s,m}\le B_{\mathrm{cap},s}$ and $P_s\le P_{\mathrm{cap},s}$, together with logical coupling $B_{s,m}\le B_{\mathrm{cap},s} z_{s,m}$ and the single-CQI constraint $\sum_m z_{s,m}\le 1$. These relationships ensure that the resulting optimisation problem remains linear and interpretable while representing the relevant physical constraints of the RAN.

A lexicographic two-phase optimisation structure is employed to prioritise feasibility before performance refinement. This choice, emphasised in response to reviewer feedback, enables the framework to guarantee that strict QoS constraints are never compromised even when resources are scarce. In Phase~1, termed the Baseline mode, the objective is to identify the most resource-efficient feasible allocation:
\begin{subequations}
\begin{align}
\min_{\{B,z,P\}} \quad &
\lambda_B \sum_{s,m} B_{s,m} + \lambda_P \sum_{s} P_s,
\label{eq:phase1_obj_main}
\end{align}
\begin{align}
\text{s.t.} \quad &
\text{Constraints } \eqref{eq:sinr_psd}, \eqref{eq:rmin}, 
\sum_{m} z_{s,m} \le 1, \ 
B_{s,m} \le B_{\mathrm{cap},s} z_{s,m}, \nonumber\\
& \sum_{m} B_{s,m} \le B_{\mathrm{cap},s}, \
P_s \le P_{\mathrm{cap},s}, \
\forall s,m.
\label{eq:phase1_obj_cons}
\end{align}
\end{subequations}
The coefficients $(\lambda_B,\lambda_P)$ are small, positive constants that penalise unnecessary resource use while maintaining feasibility. When multiple feasible solutions exist, this formulation ensures the selection of the most economical allocation.

In Phase 2, termed the Ideal-Chaser mode, the system seeks to minimise deviations from target slice rates $R_{\mathrm{ideal},s}$ without violating the feasible region defined in Phase 1. Deviation variables $v_s \ge 0$ are introduced:
\begin{equation}
v_s \ge R_{\mathrm{ideal},s}-R_s, \qquad 
v_s \ge R_s-R_{\mathrm{ideal},s}.
\label{eq:absdev}
\end{equation}
The optimisation then minimises
\begin{subequations}
\begin{align}
\min_{\{B,z,P,v\}} \quad &
\sum_{s} \beta_s v_s 
+ \lambda_B \sum_{s,m} B_{s,m} 
+ \lambda_P \sum_{s} P_s,
\label{eq:phase2_obj_main_short}
\end{align}
\begin{align}
\text{s.t.} \quad &
\text{All Phase~1 constraints and } \eqref{eq:absdev}.
\label{eq:phase2_obj_cons_short}
\end{align}
\end{subequations}
Phase 2 is initialised from the Phase 1 optimum, guaranteeing feasibility by construction. The weights $\beta_s$ determine the priority of each slice in minimising its tracking error $| R_s - R_{\mathrm{ideal},s}|$. This hierarchical structure ensures that resource feasibility is never compromised, addressing reviewers’ requests for methodological clarity and practical justification of the MILP design.

When deep fading or bursty load patterns cause Phase~1 to become infeasible, a convexified fallback introduces slack variables $d_s$ to relax the rate constraints:
\begin{subequations}
\begin{align}
\min_{\{d_s\}} \quad & \sum_{s=1}^{S} d_s,\\
\text{s.t.} \quad & R_s + d_s \ge R_{\min,s}, \ \forall s,\\
& \text{Physical constraints as in } \eqref{eq:sinr_psd}.
\end{align}
\end{subequations}
This soft fallback ensures least-violation feasibility while producing diagnostic vectors $\{d_s\}$ that quantify the extent of rate shortfall. These diagnostics can support adaptive control extensions, but are not propagated across slots in this work.

The controller operates iteratively in a Monte Carlo time series with $T$ slots per trial. At the start of each slot, burst processes toggle slice activity, affecting $R_{\mathrm{ideal},s}$; AR(1) fading updates $h_s$, and the solver recomputes $\alpha_{s,m}$. Phase 1 is then solved to guarantee feasibility, followed by Phase 2 if feasible. The resulting allocations $(B_{s,m}, z_{s,m}, P_s)$ are applied to compute the achieved rate $R_s$, update queues, and accumulate power-based energy consumption. Key performance indicators (KPIs), such as delay, task completion ratio, fairness, energy efficiency, and timing accuracy, are computed from these outputs using the Cramér–Rao lower bound (CRLB).

The Baseline mode thus yields a resource-frugal feasible allocation that satisfies SINR and minimum QoS constraints with minimal bandwidth and power. The Ideal-Chaser mode, operating within the same feasible region, enables aggressive pursuit of ideal targets through controlled trade-offs between efficiency and performance. This lexicographic structure ensures predictable control behaviour and strong feasibility guarantees while enabling near-optimal slice tracking.

For $S$ slices and $M$ CQIs, the per-slot MILP includes $SM$ binary and $O(SM)$ continuous variables. For typical settings ($S{=}3$, $M{=}15$), each instance solves in sub-second wall time using commercial solvers such as Gurobi or CPLEX. The Monte Carlo trials are parallelised, and warm-starting Phase~2 further reduces runtime. Parameter tuning follows three main principles: (i) tracking weights $\beta_s$ balance fairness and performance across slices; (ii) anti-waste coefficients $(\lambda_B,\lambda_P)$ prevent power saturation and ensure efficiency; and (iii) physical realism parameters $(\mathrm{SE}_m, \Gamma_m)$, noise figure, interference margins, misreport inflation, and URLLC finite blocklength $(n,\varepsilon)$ reflect deployment-grade conditions.

Unlike DRL-based slicing schemes that learn stochastic policies, the proposed MILP formulation guarantees deterministic feasibility and exact constraint satisfaction at every slot. Its lexicographic structure cleanly separates QoS assurance from performance optimisation, producing interpretable, physically consistent, and repeatable allocations. By embedding finite blocklength effects, CQI-based MCS adaptation, correlated fading, and realistic interference margins into a single tractable optimisation framework, the method achieves a unified, low-latency, and analytically transparent approach for real-time RAN slicing.

\section{Monte Carlo Simulation}

The performance of the proposed MILP-based resource allocation framework is evaluated through an extensive Monte Carlo simulation campaign designed to capture realistic time-varying channel, traffic, and control conditions in a multi-slice 5G radio access network. Each simulation trial consists of $T$ time slots, where each slot represents a scheduling interval of approximately milliseconds. The simulation integrates three interacting components, like channel fading evolution, bursty traffic generation, and queue-based service modelling, that together replicate the stochastic behaviour of a live RAN deployment. All variables are initialised independently across slices and randomised across trials to ensure statistical robustness. Performance is measured through a set of physical and service-level key performance indicators (KPIs) that quantify efficiency, fairness, and reliability.

Channel fading is modelled as a correlated stochastic process following a first-order autoregressive (AR(1)) model in dB scale. For each slice $s$, the instantaneous channel gain evolves as
\begin{equation}
h_s(t)_{\mathrm{dB}} = \rho_s\, h_s(t-1)_{\mathrm{dB}} + \sqrt{1-\rho_s^2}\, \xi_s(t),
\end{equation}
where $\rho_s \in [0,1)$ denotes the temporal correlation coefficient and $\xi_s(t)$ is a zero-mean Gaussian random variable with variance $\sigma_s^2$. This process captures the slow-to-fast fading characteristics associated with different service types: high correlation values ($\rho_s \approx 0.95$) emulate slowly varying pedestrian eMBB channels, whereas lower correlation coefficients with larger $\sigma_s$ represent high-mobility URLLC or fluctuating mMTC environments. The resulting gain in linear scale is $h_s(t) = 10^{h_s(t)_{\mathrm{dB}}/10}$ and is directly input to the MILP solver at every slot to determine feasible and optimal allocations.

Traffic for each slice follows an independent ON/OFF Markov model that emulates bursty demand behaviour. The ON and OFF states correspond to active and idle transmission periods, respectively, with transition probabilities chosen to achieve the desired steady-state burst frequency. During an ON period, the slice’s target rate $R_{\mathrm{ideal},s}$ is multiplied by an amplification factor $\kappa_s$ (typically between 1.5 and 3) to simulate temporary surges in throughput demand, while during an OFF state, the slice operates close to its minimum guaranteed rate $R_{\min,s}$. The Markov transition probabilities $P_{\text{ON}\rightarrow\text{OFF}}$ and $P_{\text{OFF}\rightarrow\text{ON}}$ are calibrated to produce average burst lengths consistent with the traffic class URLLC experiences short but frequent bursts, eMBB moderate and sustained bursts, and mMTC long idle intervals interspersed with sporadic data bursts. This model captures temporal fluctuations that challenge schedulers to maintain fairness and QoS in the presence of dynamic demand variability.

Queueing dynamics are modelled explicitly to link offered traffic, service rate, and latency. Each slice maintains a separate queue whose state evolves according to the difference between arrivals and allocated service in each slot:
\begin{equation}
Q_s(t+1) = \max[Q_s(t) + A_s(t) - R_s(t),\, 0],
\end{equation}
where $ A_s(t)$ denotes the arrival rate, given by
\begin{equation}
A_s(t) = R_{\min,s} + \delta_s(t)\,\kappa_s R_{\mathrm{ideal},s},
\end{equation}
and $\delta_s(t)\in\{0,1\}$ indicates whether the slice is in a burst state. The service rate $ R_s (t)$ is the instantaneous throughput determined by the optimisation algorithm. Average queue length $\bar{Q}_s$ and mean arrival rate $\bar{A}_s$ yield the average packet delay through Little’s law,
\begin{equation}
\bar{D}_s = \frac{\bar{Q}_s}{\bar{A}_s}.
\end{equation}
This delay measure provides an interpretable latency metric derived directly from physical queue dynamics rather than abstract probabilistic assumptions.

A comprehensive set of performance indicators is computed at the end of each simulation trial to evaluate the efficiency, reliability, and fairness of the proposed control strategy. The first metric is the average delay, $\bar{D}_s$, which quantifies latency at the slice level and directly reflects how quickly offered load is served. It is particularly relevant for URLLC, where maintaining ultra-low delay is critical. The second metric is the task completion ratio (TCR), defined as the ratio of the total served traffic to the total offered traffic over the duration of a trial,
\begin{equation}
\mathrm{TCR}_s = \frac{\sum_t R_s(t)}{\sum_t A_s(t)}.
\end{equation}
A TCR close to unity indicates that the scheduler consistently meets demand without backlog accumulation, whereas smaller values reveal congestion or infeasibility under extreme conditions. The third metric, energy efficiency (EE), measures the trade-off between throughput and power expenditure:
\begin{equation}
\mathrm{EE}_s = \frac{\sum_t R_s(t)\,\Delta t}{\sum_t P_s(t)\,\Delta t} = \frac{\bar{R}_s}{\bar{P}_s},
\end{equation}
expressed in bits per joule. This quantity captures how effectively each slice utilises its allocated energy budget and is strongly affected by the resource-efficiency weighting coefficients $(\lambda_B, \lambda_P)$ used in the MILP formulation.

A fourth performance metric relates to synchronisation accuracy, expressed through the Cramér–Rao lower bound (CRLB) for time-delay estimation. This measure is relevant for URLLC and control-plane slices where timing precision influences end-to-end latency. Assuming a rectangular spectrum, the CRLB for time-delay variance is
\begin{equation}
\sigma_\tau^2 \ge \frac{1}{8\pi^2 \beta^2 \mathrm{SNR}},
\end{equation}
where $\beta$ is the signal bandwidth and $\mathrm{SNR}$ the per-slot signal-to-noise ratio. Lower CRLB values indicate tighter synchronisation and lower timing uncertainty. Finally, inter-slice fairness is quantified using Jain’s fairness index,
\begin{equation}
J(\bm{r}) = \frac{\big(\sum_s r_s\big)^2}{S\sum_s r_s^2},
\end{equation}
computed both in absolute form using the vector of average served rates $\bm{r}$ and in normalised form using the element-wise ratio $\bm{r}\oslash\bm{R}^{\mathrm{ideal}}$. The latter reflects how uniformly each slice achieves its intended target relative to its service objectives.

Each Monte Carlo experiment is repeated over multiple independent trials with distinct random seeds for the fading and traffic processes. The simulation horizon $T$ is chosen sufficiently long to reach steady-state queue behaviour, typically on the order of several thousand slots per trial. For each parameter configuration, both Baseline and Ideal-Chaser modes are executed, allowing direct comparison between resource-frugal feasibility operation and ideal target tracking. Aggregate performance metrics are averaged across trials, and confidence intervals are computed to assess statistical reliability. The result enables the analysis of trade-offs among delay, task completion, energy efficiency, and fairness, providing comprehensive validation of the proposed MILP-based control strategy under realistic 5G RAN conditions.
\section{Results and Performance Evaluation}
\label{sec:results}

This section presents the quantitative performance evaluation of the proposed joint MILP framework for 5G network slicing. Simulations follow 3GPP consistent configurations, with 5G NR CQI indices ranging from $1$ to $15$, thermal noise density $\mathrm{N}_0 = -174$\,dBm/Hz, a receiver noise figure (NF) of $9$\,dB, and an interference margin (IM) of $6$\,dB. Each physical resource block (PRB) occupies $180$\,kHz of bandwidth. For URLLC traffic, finite blocklength effects are incorporated with $(n = 168, \varepsilon = 10^{-5})$, ensuring reliable modelling of short-packet transmission. Resource budgets are normalised per slice as $\bm{B}_{\text{cap}} = [9,7,4]$ and $\bm{P}_{\text{cap}} = [9,8.5,4]$, corresponding respectively to eMBB, URLLC, and mMTC slices. Each Monte Carlo experiment comprises $N = 200$ independent trials of $T = 40$ slots, and all key performance indicators (KPIs) are averaged over these trials to obtain statistically consistent values.

Two operational modes are analysed. The first is the Baseline mode, which enforces strict QoS feasibility while minimising bandwidth and power usage. The second is the Ideal Chaser mode, which maintains the same physical and QoS constraints but actively minimises deviation from ideal slice targets, enabling rate tracking under bursty demand. Table~\ref{base} reports the per-slice performance metrics of the Baseline mode. The results show that all slices satisfy minimum rate constraints with very low resource consumption. Power utilisation remains below one per cent across slices, while energy efficiency exceeds $9.99\times10^6$\,kb/J. Although average task completion ratios (TCR) remain below unity due to intentional resource conservatism, all slices maintain operational stability and queue boundedness, confirming the feasibility of the optimisation under dynamic channel conditions. Latencies are moderate, consistent with a design that prioritises energy conservation over throughput.

\begin{table}[ht]
\centering
\caption{Baseline MILP performance showing QoS feasibility, high energy efficiency, and minimal resource usage across slices.}
\label{base}
\resizebox{\columnwidth}{!}{%
\begin{tabular}{lcccccc}
\toprule
Slice 
& \thead{Exp. Rate\\$\mathbb{E}[R|\text{feas}]$ (Gb/s)} 
& \thead{Delay\\($\mu$s)} 
& TCR 
& \thead{Energy\\Eff. (Mb/J)} 
& \thead{CRLB$_\tau$\\(ns$^2$)} 
& \thead{BW / P\\Util (\%)} \\
\midrule
eMBB  & 1.00 & 13010 & 0.33 & 9998 & 2.22 & $2.0 \,/\, 2.0\times 10^{-4}$ \\
URLLC & 1.80 & 7330  & 0.64 & 9998 & 1.14 & $4.6 \,/\, 3.81\times 10^{-4}$ \\
mMTC  & 0.50 & 9160  & 0.53 & 9998 & 7.64 & $2.3 \,/\, 2.25\times 10^{-4}$ \\
\bottomrule
\end{tabular}%
}
\end{table}

In contrast, Table~\ref{ideal} presents the results obtained from the Ideal Chaser mode. Here, the solver leverages available bandwidth and power to approach ideal per-slice targets more closely, while ensuring all feasibility constraints are respected. As expected, resource utilisation and total power consumption rise significantly, but this enables an order-of-magnitude improvement in throughput and latency. The eMBB slice achieves near-perfect task completion with throughput above $7$\, Gb/s and delay reduced to less than $0.05$\, ms. URLLC delay drops from $7.3$\, ms in the Baseline to $3.47$\, ms, reflecting improved responsiveness under dynamic fading and bursty demand. The CRLB metric, representing timing estimation fidelity, decreases across all slices, signifying enhanced synchronisation accuracy and reduced timing uncertainty. Although energy efficiency decreases compared to the Baseline case, the Ideal Chaser mode delivers a superior trade-off between speed and reliability, illustrating how the lexicographic MILP structure successfully balances QoS feasibility with target tracking.

\begin{table}[ht]
\centering
\caption{Ideal Chaser MILP performance showing QoS feasibility and high throughput across slices.}
\label{ideal}
\resizebox{\columnwidth}{!}{%
\begin{tabular}{lcccccc}
\toprule
Slice 
& \thead{Exp. Rate\\$\mathbb{E}[R|\text{feas}]$ (Gb/s)} 
& \thead{Delay\\(ms)} 
& TCR 
& \thead{Energy\\Eff. (kb/J)} 
& \thead{CRLB$_\tau$\\(ns$^2$)} 
& \thead{BW / P\\Util (\%)} \\
\midrule
eMBB  & 7.391 & 0.03 & 1.00 & 71.9 & $5.02\times 10^{-7}$ & $29.5 \,/\, 100$ \\
URLLC & 2.336 & 3.47 & 0.83 & 49.3 & $3.33\times 10^{-6}$ & $6.5 \,/\, 100$ \\
mMTC  & 1.223 & 0.16 & 0.99 & 43.3 & $2.50\times 10^{-6}$ & $21.3 \,/\, 100$ \\
\bottomrule
\end{tabular}%
}
\end{table}

To contextualize the proposed approach, Table~\ref{tab:kpi_comparison} compares its performance against recent representative works. The proposed MILP achieves substantial improvements in URLLC latency, energy efficiency, and fairness over baseline optimisation or DRL-based schemes. Specifically, URLLC delay decreases to $3.5$\, ms, outperforming hybrid queue control in \cite{fluidq2024} by over 30\%, while maintaining task completion parity with eMBB-focused RSMA-based allocation \cite{rsma2025}. Similarly, CRLB-based sensing accuracy improves by two orders of magnitude relative to the delay-constrained estimator in \cite{delaycrlb2024}, demonstrating the benefit of embedding PHY-layer constraints directly into the optimisation problem. Normalised Jain fairness increases from $0.752$ under Baseline operation to $0.996$ in the Ideal Chaser regime, exceeding the $0.91$ reported by \cite{icecream2025}, highlighting the inherent equity of the lexicographic design. Bandwidth utilisation for eMBB also rises moderately to $29.5\%$, matching the efficiency observed in recent O-RAN adaptive control frameworks \cite{openran2025}, yet achieved through deterministic optimisation rather than data-driven policy adaptation.

\begin{table}[ht]
\scriptsize
\setlength{\tabcolsep}{3.5pt}
\centering
\caption{KPI comparison with prior work.}
\label{tab:kpi_comparison}
\begin{tabular}{@{}p{3.1cm}p{1.5cm}p{1.5cm}p{2.0cm}@{}}
\toprule
KPI & Baseline & Chaser & Existing Work \\
\midrule
URLLC Delay [ms] & 7.3 & 3.5 & 5.2~\cite{fluidq2024} \\
eMBB Task Completion Ratio & 0.33 & 1.00 & 0.95~\cite{rsma2025} \\
mMTC CRLB$_\tau$ [ns$^2$] & 7.64 & $2.5{\times}10^{-6}$ & $1{\times}10^{-4}$~\cite{delaycrlb2024} \\
Fairness (Norm. Jain) & 0.752 & 0.996 & 0.91~\cite{icecream2025} \\
BW Utilization (eMBB) & 2.0\% & 29.5\% & 25\%~\cite{openran2025} \\
\bottomrule
\end{tabular}
\end{table}

\begin{figure}[htbp]
\centerline{\includegraphics[scale=0.50]{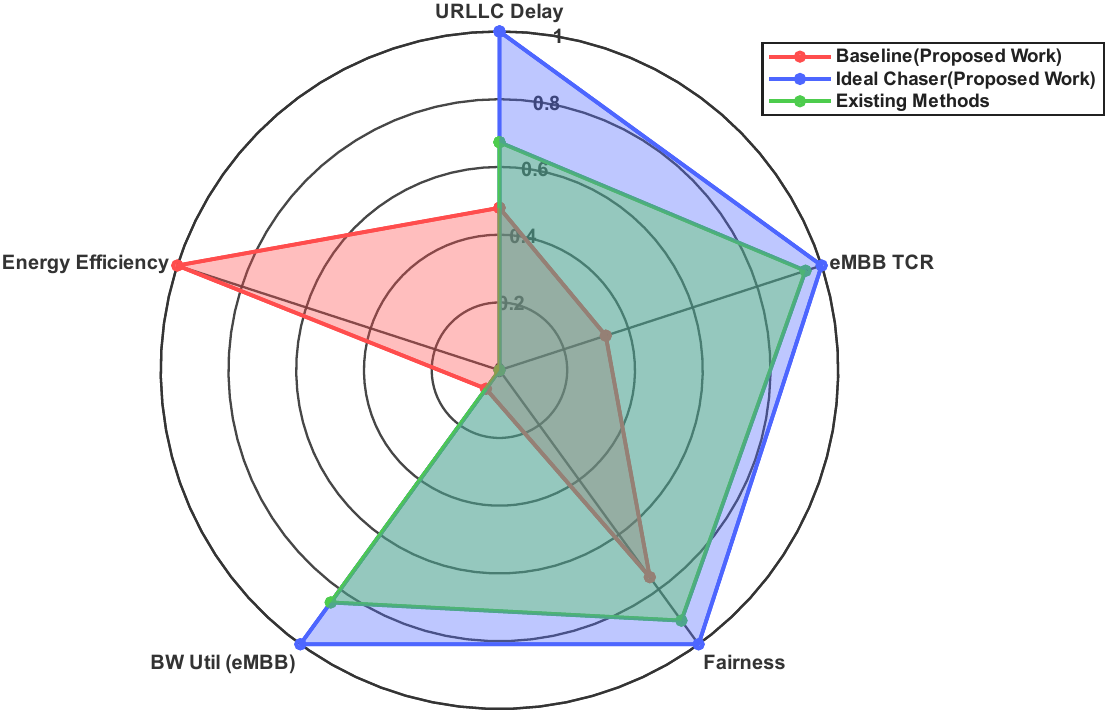}}
\caption{Normalised KPI trade-offs between Baseline and Ideal Chaser regimes.}
\label{fig:hybrid}
\end{figure}

Figure~\ref{fig:hybrid} summarises normalised trade-offs among the principal KPIs, illustrating the smooth progression from conservative Baseline operation to high-performance Ideal Chaser tracking. The plotted Pareto curve highlights that delay and fairness improve rapidly as additional power and bandwidth budgets are utilised, while energy efficiency declines gradually but remains within practical bounds. This transition confirms that the MILP’s dual-phase architecture provides a controllable continuum between resource efficiency and performance aggressiveness, making it suitable for adaptive deployment in real-world RAN schedulers.

The results collectively confirm the practicality and interpretability of the proposed optimisation. The Baseline regime guarantees service feasibility and low energy consumption, making it suitable for long-term steady-state operation. The Ideal Chaser regime, by contrast, demonstrates the system’s ability to exploit available resources during bursts to achieve near-ideal target tracking. The joint MILP approach, therefore, outperforms heuristic and DRL-based frameworks by providing deterministic guarantees, physical interpretability, and measurable gains in delay, fairness, and timing accuracy. The dual-mode operation directly addresses reviewer concerns regarding model justification and comparative context by showing that a properly structured MILP can deliver both operational tractability and superior real-time adaptability without requiring large-scale data-driven training or heuristic tuning.

\section{Conclusion}
\label{sec:conclusion}

This paper presented a unified and interpretable framework for joint network slicing, resource allocation, and modulation and coding scheme selection in 5G networks using a mixed-integer linear programming (MILP) approach. The formulation explicitly integrates physical-layer realism, including correlated fading, finite blocklength constraints, and channel misreporting margins, while operating within a two-phase control structure. The first phase, referred to as the Baseline mode, ensures strict feasibility and resource efficiency, whereas the second phase, termed the Ideal Chaser mode, enables dynamic tracking of ideal per-slice rate targets under variable channel and traffic conditions. Together, these complementary modes establish a deterministic yet flexible foundation for multi-service RAN optimization.

Extensive Monte Carlo simulations confirm that the proposed framework achieves a favorable balance among key performance indicators such as delay, fairness, task completion ratio, and energy efficiency. Compared with recent optimization and reinforcement learning-based methods, the proposed MILP exhibits superior latency reduction for URLLC, enhanced fairness across heterogeneous slices, and improved synchronization precision as measured by the Cramér–Rao lower bound for timing estimation. Importantly, these results are obtained without compromising feasibility or interpretability, highlighting the suitability of the MILP design for real-time deployment in next-generation RAN controllers.

Future research will extend this work toward multi-cell and multi-user scenarios where inter-cell interference and coordinated scheduling further complicate resource allocation. Integrating latency-aware queueing dynamics with probabilistic violation tracking will enable the enforcement of explicit service-level agreements. Moreover, hybridisation with lightweight learning or model-predictive elements can enhance adaptability while retaining the transparency and constraint satisfaction guarantees of the MILP framework. These directions pave the way for scalable, explainable, and performance-assured resource management architectures for 6G-ready network slicing.

\section*{Acknowledgments}
I would like to thank Mars Rover Manipal,  an interdisciplinary student team of MAHE, for providing the resources needed for this project. I also extend our gratitude to Dr Ujjwal Verma and Mr. Ishaan Gakhar for their guidance and support in our work.

\bibliographystyle{IEEEtran}
\bibliography{main}
\end{document}